\documentclass[acmsmall,screen,nonacm]{acmart}

\setcopyright{cc}

\usepackage{colortbl}

\usepackage{enumitem}

\begin{document}

\title{Causal Models in Requirement Specifications for Machine Learning: A vision}

\author{Hans-Martin Heyn}
\email{hans-martin.heyn@gu.se}
\affiliation{%
  \institution{Chalmers University of Technology and University of Gothenburg}
  \city{Göteborg}
  \country{Sweden}
}

\author{Yufei Mao}
\author{Roland Weiß}
\email{{yufei.mao,rolandweiss}@siemens.com}
\affiliation{%
  \institution{Siemens AG}
  \city{München}
  \country{Germany}
}

\author{Eric Knauss}
\email{Eric.Knauss@cse.gu.se}
\affiliation{%
  \institution{Chalmers University of Technology and University of Gothenburg}
  \city{Göteborg}
  \country{Sweden}
}
\renewcommand{\shortauthors}{Heyn et al.}

\begin{abstract}
Specifying data requirements for machine learning (ML) software systems remains a challenge in requirements engineering (RE). This vision paper explores causal modelling as an RE activity that allows the systematic integration of prior domain knowledge into the design of ML software systems. We propose a workflow to elicit low-level model and data requirements from high-level prior knowledge using causal models. The approach is demonstrated on an industrial fault detection system. This paper outlines future research needed to establish causal modelling as an RE practice.
\end{abstract}



\maketitle

\section{Introduction}
Rahimi et al. called for more attention towards the ability of specifying software with machine learning (ML) components~\cite{Rahimi2019}. 
Many industrial applications require \emph{robustness} of ML models against changes in input data distribution~\cite{Borg2019}. 
A key reason for lacking robustness is the difficulty of specifying ML models, because ``if input and/or output data are high-dimensional, both defining preconditions and detailed function specifications are difficult"\cite{Kuwajima2020}. 
Robustness against context changes can only be tested if the expected operational context is explicitly defined, for instance through contextual requirements\cite{Knauss2014, Knauss2016}. 
However, assumptions about the operational context are often implicit in the design process~\cite{Mitchell2021}, such as in the selection of the training dataset.
Recent surveys on requirements engineering (RE) confirm that specifying training data for ML models remains an open challenge~\cite{pei2022, ahmad2023, franch2023}. 
Current RE techniques struggle to translate high-level functional and non-functional requirements into data requirements~\cite{alves2023, villamizar2024}. 
This leads to an \emph{underspecification} causing variability in implementation choices and a lack of robustness against context changes~\cite{Fantechi2018}.\par
A possible way to address underspecification is reasoning about expected causal relationships in the ML system’s operational context. 
Typically, ML cannot infer causality from data alone~\cite{Pearl2019}. 
An ML model learns a probabilistic representation from data that seems compatible in a training context, but its performance might deviate drastically in a different operational context as statistical correlations do not capture true causal mechanisms~\cite{damour2022}.
Addressing this challenge requires incorporating prior domain knowledge and causal reasoning into the design of ML systems.\par
This vision paper proposes causal modelling to communicate \emph{prior knowledge} about causal relations in the operational context. 
We argue that by formulating prior domain knowledge as causal models we can derive requirements towards data, as well as deduce rules for runtime verification. 
This will lead to causally motivated requirements specifications for software with ML.
\paragraph{Objective of this vision paper}
First, we outline our vision of integrating causal modelling as an RE activity for ML systems. 
Then, we illustrate its application in eliciting data requirements for an industrial prototype of an ML-based cooling fault-detection system for electric motors. 
Finally, we discuss a research agenda to explore the potential of causal modelling as an RE activity for ML systems. 

\section{Related Work}
The potential of using causal modelling as part of RE activities is not yet fully explored~\cite{giamattei2024}. Fischbach et al. proposed an NLP-based process to extract and structure causal relationships from natural language~\cite{fischbach2020, fischbach2021}. A tree recursive neural network (TRNN) model was trained to detect causality in natural language requirements using logical markers such as conjunctions and negations~\cite{jadallah2021}. They further developed an approach to converts extracted causal relationships into a DAG-like structure to automatically generate test cases~\cite{fischbach2023}.
Maier et al. proposed modelling cause-effect relationships as part of scenario-based testing for automotive system safety~\cite{maier2022}. Maier et al. also introduced the concept of ``CausalOps'', an industrial lifecycle framework for causal models~\cite{maier2024}.
Gren et Brentsson Svensson proposed Bayesian Data Analysis (BDA) to evaluate the outcome of experiments on the effect of obsolete requirements on software effort estimation~\cite{gren2021}. Similarly, Frattini et al. investigated the impact of requirements quality defects on domain modelling by using BDA and causal reasoning in a in a controlled experiment~\cite{frattini2025}. While the latter two studies do not use causal modelling as an explicit RE activity, these studies demonstrates the potential of applying causal reasoning to RE activities.

\section{Causal modelling as an RE activity}
In a typical ML development pipeline, causal modelling would be a step between problem definition and data collection as it allows to formalise domain knowledge, identify relevant variables, and refine data requirements by distinguishing causal relationships from mere correlations before collecting the training data.
Particularly, graphical causal models in the form of directed acyclic graph (DAG) allow to communicate  explicitly assumed directions of causality and assumptions about \emph{confounders}, i.e., situations in which a variable $Z$ is associated to two random variables $X_1$ and $X_2$ such that a \emph{spurious relationship} between $X_1$ and $X_2$ can be observed: $X_1 \, \leftarrow \, Z \, \rightarrow \, X_2$. \par
\par

\begin{figure}[!t]
    \centering
    \includegraphics[width=0.85\linewidth]{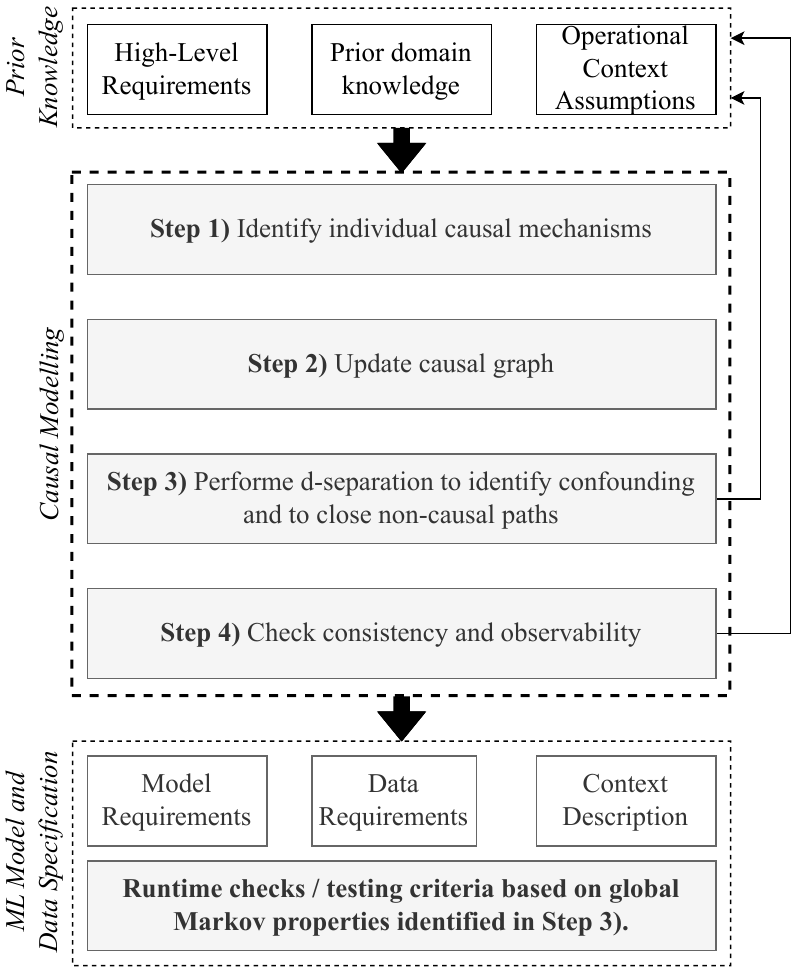}
    \caption{A proposed workflow for Causal RE}
    \label{fig:Workflow_CausalRE}
\end{figure}

Figure~\ref{fig:Workflow_CausalRE} outlines a proposed workflow. The workflow bases on the principle of \emph{causal factorisation}~\cite{Scholkopf2021}:
\begin{equation} \label{eq:decompose}
        p(X_1,\ldots,X_n) = \prod_{i=1}^{n} p\left(X_i|\mathbf{PA}_i\right), 
\end{equation} 
where $\mathbf{PA}_i$ denotes the set of parents (variables that have a direct causal effect) of a variable $X_i$ in the DAG.
\emph{Causal factorisation} implies that an observed joint distribution of interest can be decomposed into a product of conditional distributions, where each term corresponds to a causal mechanism. \par
\textbf{Step 1) Identify individual causal mechanisms:} The aim is to pinpoint specific cause-effect pathways informed based on high-level requirements, prior domain knowledge, and context assumptions.\par
\textbf{Step 2) Update causal graph:} Once a causal mechanism is identified, the relevant observable and latent variables are determined, and a causal graph is updated to include these variables along with the assumed directions of cause-and-effect relationships. \par



\textbf{Step 3) Perform d-separation and extract requirements:} 
With the causal model, \emph{d-separation}\footnote{Due to space constraints, background on d-separation is omitted but can be found in~\cite{Pearl2019}.} allows to identify variables that are needed to block ``non-causal'' association paths.  
Taking the example from above, in $X_1 \leftarrow Z \rightarrow X_2$, there is a ``non-causal'' path between $X_1$ and $X_2$.  
If the ML model can condition on $Z$ (assuming $Z$ is observable), $X_1$ and $X_2$ become d-separated, closing the ``non-causal'' path.
This is an example of a resulting data requirement: 
$Z$ must be included in the training dataset to avoid learning a spurious correlation between $X_1$ and $X_2$.  
Additionally, Step 3 provides \emph{independence criteria} based on global Markov properties:  
If $X_1$ and $X_2$ are d-separated by $Z$, they are conditionally independent given $Z$, i.e., $X_1 \perp X_2 \, |\, Z$.  
This provides \emph{testable criteria} to verify prior knowledge and assumptions encoded in the causal graph.\par

\textbf{Step 4) Check consistency and observability:}
The graphical causal model must be checked for cyclic dependencies because a variable cannot be its own cause~\cite{Glymour2016}.  
Furthermore, variables needed to block ``non-causal'' paths must be observable.  
If this is not the case, the system must be adjusted to enable their observation or suitable instrument variables must be identified~\cite{angrist1996}.\par
The resulting causal graph becomes part of an ML specification because it encodes the assumed causal structures, prior knowledge, and operational context, from which data and model requirements, as well as testing criteria, are derived.  

\section{Demonstration on industrial prototype}
We demonstrate the use of causal modelling as an RE activity on an industrial prototype use case, specifically a system for detecting faults in the cooling system of electric motors.

\paragraph{Methodology:}
We held three workshops with two Siemens engineers and two academic researchers to explore using causal models for requirements specification in the second half of 2022.
The researchers introduced causal models with examples like \emph{temperature} $\leftarrow$ \emph{sunrise} $\rightarrow$ \emph{birds chirping} and explaining key concepts such as \emph{confounding}, \emph{colliders}, and \emph{d-separation} using for example the \emph{back-door criteria}.  
The company experts then presented the prototype system, and prior knowledge rules were formalised together by identifying causal mechanisms and updating the causal model iteratively with each newly found causal mechanism. 
We then applied \emph{d-separation} to close non-causal paths between the exposure (i.e., a cooling fault) and the outcome (i.e., the classification result) which resulted in data and model requirements to ensure the ML model controls for potential confounding.

\par
\begin{figure}[!ht]
    \centering
    \includegraphics[width=\linewidth]{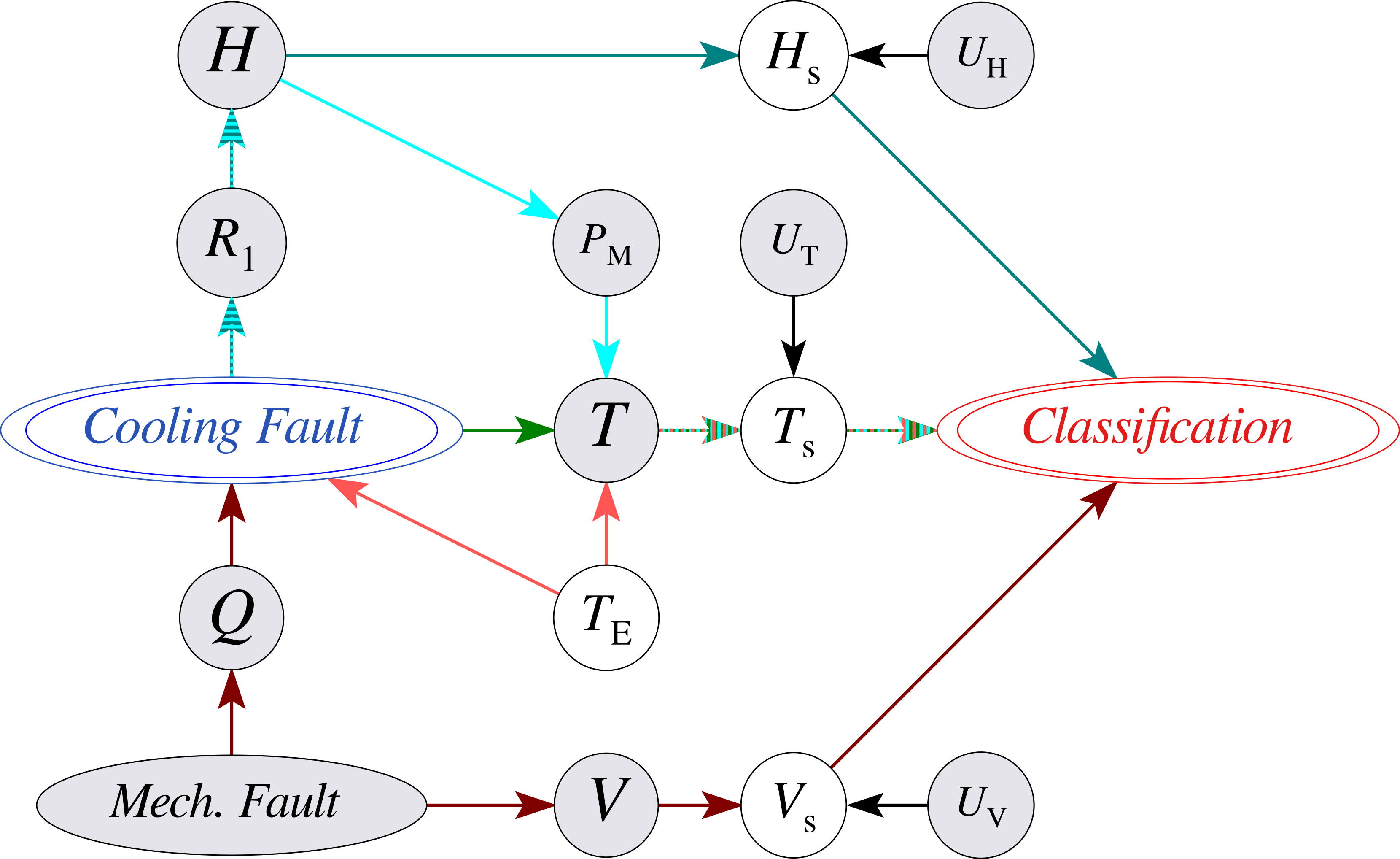}
    \caption{DAG for the motor diagnostic use case. Gray-background nodes are latent (unobservable) variables, while white-background nodes are observable at runtime.}
    \label{fig:motor_diagnostic_DAG}
\end{figure}

\paragraph{Description of demonstration case:}
The demonstration case, provided by Siemens, is a motor diagnostic application for monitoring electrical motors  using an attachable sensor device. Initially, the system detected cooling faults from vibrations caused by mechanical faults, such as a broken fan blade. The new device will use an ML model to detect faults based on multiple sensor inputs. The high-level functional requirement is:  

\begin{description}
    \item[FR-1:] \emph{GIVEN indoor operational environment WHEN the cooling system is abnormal THEN an alarm should be raised.}
\end{description}
The following prior knowledge of the company engineers was considered for identifying causal mechanisms:
\begin{description}
    \item[PK-1:] A fault in the cooling system can affect the magnetic flux by changing the temperature of the rotor material and thus affecting the electrical resistance.
    \item[PK-2:] Mechanical faults of the fan can reduce the available airflow.
    \item[PK-3:] Mechanical faults cause vibrations of the system.
    \item[PK-4:] Environmental temperature has an influence on the temperature signal because the sensor is mounted outside the motor. 
    \item[PK-5:] Unmeasured sensor disturbances exist in general.
\end{description}
\paragraph{Results:}
The resulting causal model for the motor diagnostic use case is shown in Figure~\ref{fig:motor_diagnostic_DAG}. 

Explanations for the variables and their relations to the prior knowledge are provided in Table~\ref{tab:motorcond}. 

\begin{table}[!t]
\centering
\caption{Variables for motor diagnostic use case.} 
\label{tab:motorcond}
\begin{tabular}{l c l}
\textbf{Variable}    & \textbf{Related PKs} & \textbf{Definition}               \\ \midrule
Cooling Fault    & PK1, PK2   & Fan system status         \\ 
$Q$         & PK2       & Max. possible airflow           \\ 
Mech. Fault & PK2, PK3 & Mechanical fault of motor \\ 
$P_M$       & PK1         & Mechanical power          \\ 
$R_1$       & PK1         & Electrical (inner) losses   \\ 
$T_E$       & PK4         & Environmental temperature              \\ 
$U_X$       & PK5         & Unmeasured noises              \\ 
$T$ ($T_s$) &               & Surface temperature (measured) \\ 
$H$ ($H_s$) &               & Magnetic Flux (measured)               \\ 
$V$ ($V_s$) &               & Vibrations (measured)        \\ \bottomrule
\end{tabular}
\end{table}

The causal graph in Figure~\ref{fig:motor_diagnostic_DAG} includes three causal mechanisms between \emph{Cooling Fault} occurrence and \emph{Classification} whether or not a cooling fault has occurred: \par 
\definecolor{GreenHTML}{HTML}{008000}
\textbf{\textcolor{GreenHTML}{Temperature mechanism}:} A cooling fault increases the motor's surface temperature $T$ (via the core temperature), measured by the temperature sensor $T_s$, which can be used to classify a cooling fault.

\definecolor{BlueHTML}{HTML}{008080}
\textbf{\textcolor{BlueHTML}{Magnetic flux mechanism}:} A cooling fault changes the inner resistance (via the core temperature), which affect the magnetic flux $H$. This is measured by the fluxmeter $H_s$ for classification.  

\definecolor{LightBlueHTML}{HTML}{00ffff}
\textbf{\textcolor{LightBlueHTML}{Mechanical power mechanism}:} A cooling fault changes the magnetic flux $H$, which affects the mechanical power $P_M$ and surface temperature $T$. The latter is measured by the sensor $T_s$ for classification. \par \pagebreak
\noindent Two confounding paths were identified:\par
\definecolor{RedHTML}{HTML}{800000}
\textbf{\textcolor{RedHTML}{Mechanical fault confounding}:} A mechanical fan blade fault can reduce the available airflow $Q$ causing a cooling fault and vibrations $V$, which are measured by a vibration sensor $V_s$ for classification.  

\definecolor{BrightRedHTML}{HTML}{ff5555}
\textbf{\textcolor{BrightRedHTML}{Environmental temperature confounding}:} A sudden change in environment temperature $T_E$ can temporarily limit cooling without indicating a fault and it affects the surface temperature $T$.\par

\paragraph{Data and model requirements:} 
We checked which variables must be observed and controlled for to close non-causal paths between cooling fault occurrence and classification of a cooling fault, which resulted in the requirements listed in Table~\ref{tab:motorrqs}. 
Vibration data $V_s$ alone is insufficient to detect cooling faults, as not all mechanical faults lead to a cooling fault (RQ-D1, RQ-M1). Instead, data on temperature and magnetic flux mechanisms should be included (RQ-M2).\footnote{In fact, vibration data may be unnecessary for detecting cooling faults unless it is desired to distinguish mechanical from non-mechanical causes.} An additional sensor should record the environmental temperature $T_E$ to control for confounding (RQ-D2). Sensor noise must also be represented in the training data (RQ-D3). 

\begin{table}[!t]
\centering
\caption{Requirements derived from causal graph}
\label{tab:motorrqs}
\begin{tabular}{lp{12.0cm}}
\textbf{ID} & \textbf{Requirement (RQ-D: Data Req., RQ-M: Model Req.)}  \\ \midrule
RQ-D1 & Training data shall include cases where mechanical faults cause vibrations $V$ without leading to cooling faults . \\
RQ-D2 & The occurrence of cooling faults shall be conditioned on different environmental temperatures $T_E$ such that the model can learn the confounding influence of $T_E$. \\ 
RQ-D3 & Measurements shall include characteristic sensor noise. \\
RQ-M1 & Cooling faults shall not be classified based on vibration data $V_s$ alone. \\
RQ-M2 & The input layer shall accept temperature, magnetic flux, and vibration measurements. \\

\bottomrule
\end{tabular}
\end{table}

\paragraph{Testing and runtime checks:} The causal graph in Figure~\ref{fig:motor_diagnostic_DAG} implies a set of independence conditions:
\[
\begin{aligned}
    \textbf{ID1}:~&\text{Classification} \perp T_E \mid H_s, T_s, V_s \\
    \textbf{ID2}:~&H_s \perp T_E \mid \text{Cooling Fault} \\
    \textbf{ID3}:~&H_s \perp V_s \mid \text{Cooling Fault} \\
    \textbf{ID4}:~&T_s \perp V_s \mid \text{Cooling Fault}, T_E \\
    \textbf{ID5}:~&V_s \perp T_E
\end{aligned}
\]
As an example for a resulting test case, ID1 states that classification is independent of $T_E$ given $H_s$, $T_s$, and $V_s$. A test case could trigger faults at varying $T_E$ to verify that the detection probability remains unchanged. 
As an example for runtime monitoring, ID5 suggests $V_s$ and $T_E$ should be independent. An additional monitor could track their correlation during operation and trigger an alarm if a threshold is exceeded which would indicate a shift in the assumed operational context (e.g., the probability of a mechanical fault could depend on the environmental temperature which would be a violation of the assumed causal models for this system).


\section{Discussion and research agenda}
In this vision paper we argue that causal modelling and its mathematical framework have significant potential as an RE activity for ML software system development by systematically integrating prior knowledge into the design. 
However, based on the experience in our demonstration use case, further research is needed before this vision becomes standard industry practice.\par
\textit{Causal models as complement to natural language requirements.} Causal graphs originates from mathematics. We must explore how they can complement current requirements specifications and how they must be adopted for RE. Terms like ``treatment'', ``confounder'', and ``collider'' are uncommon in RE and require interpretation.\par
\textit{Criteria for sufficient variable selection.} A key challenge is knowing when a causal graphs includes ``enough'' prior knowledge. We need methods to determine a sufficient set of variables that must be included for a given use case and methods for deciding between competing causal DAGs given a ´´sufficient'' set of variables.\par
\textit{Modularisation of ML software systems.} Isolating causal mechanisms can guide the modularisation of ML systems, i.e., dividing large monolithic ML models into smaller sub-models.\par
\textit{A common language between different stakeholders of ML software systems.} Causal models provide a unified way to communicate prior knowledge and assumptions. Research should explore how this can facilitate coordination between different groups such as data scientists, product experts, and software engineers.\par
\textit{Data requirements derived through causal reasoning.} Causal reasoning in RE helps identifying data requirements. 
Further research should assess to what degree data requirements derived from causal models can enhance ML robustness and reduce data needs compared to traditional RE methods.\par
\textit{Testing and runtime checks.} 
ML software system must align with expected (causal) behaviour. Causal graphs imply independence criteria that lead to \emph{testable implications} for the runtime behaviour. Research should explore how to translate these into testing strategies and monitors and how reliable such monitors are in practice. 

\paragraph*{Conclusion}
Causal reasoning offers a systematic way to integrate prior knowledge into RE for ML software systems. 
We outlined a vision and demonstrated a preliminary workflow to derive and argue for low level model and data requirements from high level prior knowledge using causal graphs. We discussed future research activities that are needed to turn this vision into industrial practice. 

\begin{acks}
This project has received funding from the EU’s Horizon research and innovation program under grant agreement No 957197 (vedliot).
\end{acks}

\bibliographystyle{ACM-Reference-Format}
\bibliography{bib}


\begin{thebibliography}{26}


\ifx \showCODEN    \undefined \def \showCODEN     #1{\unskip}     \fi
\ifx \showDOI      \undefined \def \showDOI       #1{#1}\fi
\ifx \showISBNx    \undefined \def \showISBNx     #1{\unskip}     \fi
\ifx \showISBNxiii \undefined \def \showISBNxiii  #1{\unskip}     \fi
\ifx \showISSN     \undefined \def \showISSN      #1{\unskip}     \fi
\ifx \showLCCN     \undefined \def \showLCCN      #1{\unskip}     \fi
\ifx \shownote     \undefined \def \shownote      #1{#1}          \fi
\ifx \showarticletitle \undefined \def \showarticletitle #1{#1}   \fi
\ifx \showURL      \undefined \def \showURL       {\relax}        \fi
\providecommand\bibfield[2]{#2}
\providecommand\bibinfo[2]{#2}
\providecommand\natexlab[1]{#1}
\providecommand\showeprint[2][]{arXiv:#2}

\bibitem[Ahmad et~al\mbox{.}(2023)]%
        {ahmad2023}
\bibfield{author}{\bibinfo{person}{Khlood Ahmad}, \bibinfo{person}{Mohamed Abdelrazek}, \bibinfo{person}{Chetan Arora}, \bibinfo{person}{Muneera Bano}, {and} \bibinfo{person}{John Grundy}.} \bibinfo{year}{2023}\natexlab{}.
\newblock \showarticletitle{Requirements engineering for artificial intelligence systems: A systematic mapping study}.
\newblock \bibinfo{journal}{\emph{IST}}  \bibinfo{volume}{158} (\bibinfo{year}{2023}), \bibinfo{pages}{107176}.
\newblock


\bibitem[Alves et~al\mbox{.}(2023)]%
        {alves2023}
\bibfield{author}{\bibinfo{person}{Antonio Pedro~Santos Alves}, \bibinfo{person}{Marcos Kalinowski}, \bibinfo{person}{G{\"o}rkem Giray}, \bibinfo{person}{Daniel Mendez}, \bibinfo{person}{Niklas Lavesson}, \bibinfo{person}{Kelly Azevedo}, \bibinfo{person}{Hugo Villamizar}, \bibinfo{person}{Tatiana Escovedo}, \bibinfo{person}{Helio Lopes}, \bibinfo{person}{Stefan Biffl}, {et~al\mbox{.}}} \bibinfo{year}{2023}\natexlab{}.
\newblock \showarticletitle{Status quo and problems of requirements engineering for machine learning: Results from an international survey}. In \bibinfo{booktitle}{\emph{Int. Conf. on Product-Focused Soft. Proc. Improv.}} Springer, \bibinfo{pages}{159--174}.
\newblock


\bibitem[Angrist et~al\mbox{.}(1996)]%
        {angrist1996}
\bibfield{author}{\bibinfo{person}{Joshua~D Angrist}, \bibinfo{person}{Guido~W Imbens}, {and} \bibinfo{person}{Donald~B Rubin}.} \bibinfo{year}{1996}\natexlab{}.
\newblock \showarticletitle{Identification of causal effects using instrumental variables}.
\newblock \bibinfo{journal}{\emph{Journal of the American Stat. Asso.}} \bibinfo{volume}{91}, \bibinfo{number}{434} (\bibinfo{year}{1996}), \bibinfo{pages}{444--455}.
\newblock


\bibitem[Borg et~al\mbox{.}(2019)]%
        {Borg2019}
\bibfield{author}{\bibinfo{person}{Markus Borg}, \bibinfo{person}{Cristofer Englund}, \bibinfo{person}{Krzysztof Wnuk}, \bibinfo{person}{Boris Duran}, \bibinfo{person}{Christoffer Levandowski}, \bibinfo{person}{Shenjian Gao}, \bibinfo{person}{Yanwen Tan}, \bibinfo{person}{Henrik Kaijser}, \bibinfo{person}{Henrik L{\"o}nn}, {and} \bibinfo{person}{Jonas T{\"o}rnqvist}.} \bibinfo{year}{2019}\natexlab{}.
\newblock \showarticletitle{Safely Entering the Deep: A Review of Verification and Validation for Machine Learning and a Challenge Elicitation in the Automotive Industry}.
\newblock \bibinfo{journal}{\emph{Journal of Automotive Software Engineering}} \bibinfo{volume}{1}, \bibinfo{number}{1} (\bibinfo{year}{2019}), \bibinfo{pages}{1--19}.
\newblock


\bibitem[D'Amour et~al\mbox{.}(2022)]%
        {damour2022}
\bibfield{author}{\bibinfo{person}{Alexander D'Amour}, \bibinfo{person}{Katherine Heller}, \bibinfo{person}{Dan Moldovan}, \bibinfo{person}{Ben Adlam}, \bibinfo{person}{Babak Alipanahi}, \bibinfo{person}{Alex Beutel}, \bibinfo{person}{Christina Chen}, \bibinfo{person}{Jonathan Deaton}, \bibinfo{person}{Jacob Eisenstein}, \bibinfo{person}{Matthew~D Hoffman}, {et~al\mbox{.}}} \bibinfo{year}{2022}\natexlab{}.
\newblock \showarticletitle{Underspecification presents challenges for credibility in modern machine learning}.
\newblock \bibinfo{journal}{\emph{Journal of Machine Learning Research}} \bibinfo{volume}{23}, \bibinfo{number}{226} (\bibinfo{year}{2022}), \bibinfo{pages}{1--61}.
\newblock


\bibitem[Fantechi et~al\mbox{.}(2018)]%
        {Fantechi2018}
\bibfield{author}{\bibinfo{person}{Alessandro Fantechi}, \bibinfo{person}{Alessio Ferrari}, \bibinfo{person}{Stefania Gnesi}, {and} \bibinfo{person}{Laura Semini}.} \bibinfo{year}{2018}\natexlab{}.
\newblock \showarticletitle{Requirement engineering of software product lines: Extracting variability using NLP}. In \bibinfo{booktitle}{\emph{2018 26th IEEE RE Conf.}} IEEE, \bibinfo{pages}{418--423}.
\newblock


\bibitem[Fischbach et~al\mbox{.}(2021)]%
        {fischbach2021}
\bibfield{author}{\bibinfo{person}{Jannik Fischbach}, \bibinfo{person}{Julian Frattini}, \bibinfo{person}{Arjen Spaans}, \bibinfo{person}{Maximilian Kummeth}, \bibinfo{person}{Andreas Vogelsang}, \bibinfo{person}{Daniel Mendez}, {and} \bibinfo{person}{Michael Unterkalmsteiner}.} \bibinfo{year}{2021}\natexlab{}.
\newblock \showarticletitle{Automatic detection of causality in requirement artifacts: the cira approach}. In \bibinfo{booktitle}{\emph{2021 27th REFSQ}}. Springer, \bibinfo{pages}{19--36}.
\newblock


\bibitem[Fischbach et~al\mbox{.}(2023)]%
        {fischbach2023}
\bibfield{author}{\bibinfo{person}{Jannik Fischbach}, \bibinfo{person}{Julian Frattini}, \bibinfo{person}{Andreas Vogelsang}, \bibinfo{person}{Daniel Mendez}, \bibinfo{person}{Michael Unterkalmsteiner}, \bibinfo{person}{Andreas Wehrle}, \bibinfo{person}{Pablo~Restrepo Henao}, \bibinfo{person}{Parisa Yousefi}, \bibinfo{person}{Tedi Juricic}, \bibinfo{person}{Jeannette Radduenz}, {et~al\mbox{.}}} \bibinfo{year}{2023}\natexlab{}.
\newblock \showarticletitle{Automatic creation of acceptance tests by extracting conditionals from requirements: NLP approach and case study}.
\newblock \bibinfo{journal}{\emph{JSS}}  \bibinfo{volume}{197} (\bibinfo{year}{2023}), \bibinfo{pages}{111549}.
\newblock


\bibitem[Fischbach et~al\mbox{.}(2020)]%
        {fischbach2020}
\bibfield{author}{\bibinfo{person}{Jannik Fischbach}, \bibinfo{person}{Benedikt Hauptmann}, \bibinfo{person}{Lukas Konwitschny}, \bibinfo{person}{Dominik Spies}, {and} \bibinfo{person}{Andreas Vogelsang}.} \bibinfo{year}{2020}\natexlab{}.
\newblock \showarticletitle{Towards causality extraction from requirements}. In \bibinfo{booktitle}{\emph{2020 28th IEEE RE Conf.}} IEEE, \bibinfo{pages}{388--393}.
\newblock


\bibitem[Franch et~al\mbox{.}(2023)]%
        {franch2023}
\bibfield{author}{\bibinfo{person}{Xavier Franch}, \bibinfo{person}{Andreas Jedlitschka}, {and} \bibinfo{person}{Silverio Mart{\'\i}nez-Fern{\'a}ndez}.} \bibinfo{year}{2023}\natexlab{}.
\newblock \showarticletitle{A requirements engineering perspective to AI-based systems development: A vision paper}. In \bibinfo{booktitle}{\emph{2023 REFSQ}}. Springer, \bibinfo{pages}{223--232}.
\newblock


\bibitem[Frattini et~al\mbox{.}(2025)]%
        {frattini2025}
\bibfield{author}{\bibinfo{person}{Julian Frattini}, \bibinfo{person}{Davide Fucci}, \bibinfo{person}{Richard Torkar}, \bibinfo{person}{Lloyd Montgomery}, \bibinfo{person}{Michael Unterkalmsteiner}, \bibinfo{person}{Jannik Fischbach}, {and} \bibinfo{person}{Daniel Mendez}.} \bibinfo{year}{2025}\natexlab{}.
\newblock \showarticletitle{Applying bayesian data analysis for causal inference about requirements quality: a controlled experiment}.
\newblock \bibinfo{journal}{\emph{Empirical Software Engineering}} \bibinfo{volume}{30}, \bibinfo{number}{1} (\bibinfo{year}{2025}), \bibinfo{pages}{29}.
\newblock


\bibitem[Giamattei et~al\mbox{.}(2024)]%
        {giamattei2024}
\bibfield{author}{\bibinfo{person}{Luca Giamattei}, \bibinfo{person}{Antonio Guerriero}, \bibinfo{person}{Roberto Pietrantuono}, {and} \bibinfo{person}{Stefano Russo}.} \bibinfo{year}{2024}\natexlab{}.
\newblock \showarticletitle{Causal reasoning in Software Quality Assurance: A systematic review}.
\newblock \bibinfo{journal}{\emph{IST}} (\bibinfo{year}{2024}), \bibinfo{pages}{107599}.
\newblock


\bibitem[Glymour et~al\mbox{.}(2016)]%
        {Glymour2016}
\bibfield{author}{\bibinfo{person}{Madelyn Glymour}, \bibinfo{person}{Judea Pearl}, {and} \bibinfo{person}{Nicholas~P Jewell}.} \bibinfo{year}{2016}\natexlab{}.
\newblock \bibinfo{booktitle}{\emph{Causal inference in statistics: A primer}}.
\newblock \bibinfo{publisher}{John Wiley \& Sons}.
\newblock


\bibitem[Gren and Berntsson~Svensson(2021)]%
        {gren2021}
\bibfield{author}{\bibinfo{person}{Lucas Gren} {and} \bibinfo{person}{Richard Berntsson~Svensson}.} \bibinfo{year}{2021}\natexlab{}.
\newblock \showarticletitle{Is it possible to disregard obsolete requirements? a family of experiments in software effort estimation}.
\newblock \bibinfo{journal}{\emph{REJ}} \bibinfo{volume}{26}, \bibinfo{number}{3} (\bibinfo{year}{2021}), \bibinfo{pages}{459--480}.
\newblock


\bibitem[Jadallah et~al\mbox{.}(2021)]%
        {jadallah2021}
\bibfield{author}{\bibinfo{person}{Noah Jadallah}, \bibinfo{person}{Jannik Fischbach}, \bibinfo{person}{Julian Frattini}, {and} \bibinfo{person}{Andreas Vogelsang}.} \bibinfo{year}{2021}\natexlab{}.
\newblock \showarticletitle{Cate: Causality tree extractor from natural language requirements}. In \bibinfo{booktitle}{\emph{2021 29th RE Conf. Workshops (REW)}}. IEEE, \bibinfo{pages}{77--79}.
\newblock


\bibitem[Knauss et~al\mbox{.}(2016)]%
        {Knauss2016}
\bibfield{author}{\bibinfo{person}{Alessia Knauss}, \bibinfo{person}{Daniela Damian}, \bibinfo{person}{Xavier Franch}, \bibinfo{person}{Angela Rook}, \bibinfo{person}{Hausi~A M{\"u}ller}, {and} \bibinfo{person}{Alex Thomo}.} \bibinfo{year}{2016}\natexlab{}.
\newblock \showarticletitle{ACon: A learning-based approach to deal with uncertainty in contextual requirements at runtime}.
\newblock \bibinfo{journal}{\emph{IST}}  \bibinfo{volume}{70} (\bibinfo{year}{2016}), \bibinfo{pages}{85--99}.
\newblock


\bibitem[Knauss et~al\mbox{.}(2014)]%
        {Knauss2014}
\bibfield{author}{\bibinfo{person}{Alessia Knauss}, \bibinfo{person}{Daniela Damian}, {and} \bibinfo{person}{Kurt Schneider}.} \bibinfo{year}{2014}\natexlab{}.
\newblock \showarticletitle{Eliciting contextual requirements at design time: A case study}. In \bibinfo{booktitle}{\emph{2014 4th Int. Workshop on Emp. RE}}. IEEE, \bibinfo{pages}{56--63}.
\newblock


\bibitem[Kuwajima et~al\mbox{.}(2020)]%
        {Kuwajima2020}
\bibfield{author}{\bibinfo{person}{Hiroshi Kuwajima}, \bibinfo{person}{Hirotoshi Yasuoka}, {and} \bibinfo{person}{Toshihiro Nakae}.} \bibinfo{year}{2020}\natexlab{}.
\newblock \showarticletitle{Engineering problems in machine learning systems}.
\newblock \bibinfo{journal}{\emph{MLg}} \bibinfo{volume}{109}, \bibinfo{number}{5} (\bibinfo{year}{2020}), \bibinfo{pages}{1103--1126}.
\newblock


\bibitem[Maier et~al\mbox{.}(2022)]%
        {maier2022}
\bibfield{author}{\bibinfo{person}{Robert Maier}, \bibinfo{person}{Lisa Grabinger}, \bibinfo{person}{David Urlhart}, {and} \bibinfo{person}{J{\"u}rgen Mottok}.} \bibinfo{year}{2022}\natexlab{}.
\newblock \showarticletitle{Towards causal model-based engineering in automotive system safety}. In \bibinfo{booktitle}{\emph{Int. Symp. on Model-Based Safety}}. Springer, \bibinfo{pages}{116--129}.
\newblock


\bibitem[Maier et~al\mbox{.}(2024)]%
        {maier2024}
\bibfield{author}{\bibinfo{person}{Robert Maier}, \bibinfo{person}{Andreas Schlattl}, \bibinfo{person}{Thomas Guess}, {and} \bibinfo{person}{J{\"u}rgen Mottok}.} \bibinfo{year}{2024}\natexlab{}.
\newblock \showarticletitle{CausalOps—Towards an industrial lifecycle for causal probabilistic graphical models}.
\newblock \bibinfo{journal}{\emph{IST}}  \bibinfo{volume}{174} (\bibinfo{year}{2024}), \bibinfo{pages}{107520}.
\newblock


\bibitem[Mitchell et~al\mbox{.}(2021)]%
        {Mitchell2021}
\bibfield{author}{\bibinfo{person}{Shira Mitchell}, \bibinfo{person}{Eric Potash}, \bibinfo{person}{Solon Barocas}, \bibinfo{person}{Alexander D'Amour}, {and} \bibinfo{person}{Kristian Lum}.} \bibinfo{year}{2021}\natexlab{}.
\newblock \showarticletitle{Algorithmic fairness: Choices, assumptions, and definitions}.
\newblock \bibinfo{journal}{\emph{Ann. Review of Stat. and Its Appl.}}  \bibinfo{volume}{8} (\bibinfo{year}{2021}), \bibinfo{pages}{141--163}.
\newblock


\bibitem[Pearl(2019)]%
        {Pearl2019}
\bibfield{author}{\bibinfo{person}{Judea Pearl}.} \bibinfo{year}{2019}\natexlab{}.
\newblock \showarticletitle{{The Limitations of Opaque Learning Machines}}.
\newblock In \bibinfo{booktitle}{\emph{Possible Minds: 25 Ways of Looking at AI}}, \bibfield{editor}{\bibinfo{person}{Johnm Brockman}} (Ed.). \bibinfo{publisher}{Penguin Press}, \bibinfo{address}{London}, Chapter~2.
\newblock


\bibitem[Pei et~al\mbox{.}(2022)]%
        {pei2022}
\bibfield{author}{\bibinfo{person}{Zhongyi Pei}, \bibinfo{person}{Lin Liu}, \bibinfo{person}{Chen Wang}, {and} \bibinfo{person}{Jianmin Wang}.} \bibinfo{year}{2022}\natexlab{}.
\newblock \showarticletitle{Requirements engineering for machine learning: A review and reflection}. In \bibinfo{booktitle}{\emph{2022 30th IEEE RE Conf. Workshops (REW)}}. IEEE, \bibinfo{pages}{166--175}.
\newblock


\bibitem[Rahimi et~al\mbox{.}(2019)]%
        {Rahimi2019}
\bibfield{author}{\bibinfo{person}{Mona Rahimi}, \bibinfo{person}{Jin~LC Guo}, \bibinfo{person}{Sahar Kokaly}, {and} \bibinfo{person}{Marsha Chechik}.} \bibinfo{year}{2019}\natexlab{}.
\newblock \showarticletitle{Toward requirements specification for machine-learned components}. In \bibinfo{booktitle}{\emph{2019 27th IEEE RE Workshops (REW)}}. IEEE, \bibinfo{pages}{241--244}.
\newblock


\bibitem[Sch{\"o}lkopf et~al\mbox{.}(2021)]%
        {Scholkopf2021}
\bibfield{author}{\bibinfo{person}{Bernhard Sch{\"o}lkopf}, \bibinfo{person}{Francesco Locatello}, \bibinfo{person}{Stefan Bauer}, \bibinfo{person}{Nan~Rosemary Ke}, \bibinfo{person}{Nal Kalchbrenner}, \bibinfo{person}{Anirudh Goyal}, {and} \bibinfo{person}{Yoshua Bengio}.} \bibinfo{year}{2021}\natexlab{}.
\newblock \showarticletitle{Toward causal representation learning}.
\newblock \bibinfo{journal}{\emph{Proc. IEEE}} \bibinfo{volume}{109}, \bibinfo{number}{5} (\bibinfo{year}{2021}), \bibinfo{pages}{612--634}.
\newblock


\bibitem[Villamizar et~al\mbox{.}(2024)]%
        {villamizar2024}
\bibfield{author}{\bibinfo{person}{Hugo Villamizar}, \bibinfo{person}{Marcos Kalinowski}, \bibinfo{person}{H{\'e}lio Lopes}, {and} \bibinfo{person}{Daniel Mendez}.} \bibinfo{year}{2024}\natexlab{}.
\newblock \showarticletitle{Identifying concerns when specifying machine learning-enabled systems: a perspective-based approach}.
\newblock \bibinfo{journal}{\emph{JSS}}  \bibinfo{volume}{213} (\bibinfo{year}{2024}), \bibinfo{pages}{112053}.
\newblock


\end{thebibliography}


\end{document}